\documentstyle[11pt]{article}

\begin{document}
\begin{titlepage}
\begin{flushright}
{\small DE-FG05-92ER40717-7}
\end{flushright}
\vspace*{18mm}
\begin{center}
               {\LARGE\bf Optimization of $R_{e^+e^-}$ \\
                  and \\
\vspace*{4mm}
               ``Freezing'' of the QCD Couplant at Low Energies}
\vspace{20mm}\\
{\Large A. C. Mattingly and P. M. Stevenson}
\vspace{18mm}\\
{\large\it
T.W. Bonner Laboratory, Physics Department,\\
Rice University, Houston, TX 77251, USA}
\vspace{25mm}\\
{\bf Abstract:}
\end{center}

The new result for the third-order QCD corrections to $R_{e^+e^-}$,
unlike the old, incorrect result, is nicely compatible with the
principle-of-minimal-sensitivity optimization method.  Moreover,
it leads to infrared fixed-point behaviour:  the optimized couplant,
$\alpha_s/\pi$, for $R_{e^+e^-}$ does not diverge at low energies, but
``freezes'' to a value $0.26$ below about 300 MeV.  This provides
some direct theoretical evidence, purely from perturbation theory,
for the ``freezing'' of the couplant -- an idea that has long been
a popular and successful phenomenological hypothesis.  We use the
``smearing'' method of Poggio, Quinn, and Weinberg to compare the
resulting theoretical prediction for $R_{e^+e^-}$ with experimental
data down to the lowest energies, and find excellent agreement.

\end{titlepage}

\setcounter{page}{1}

\section{Introduction}

The calculation {\cite{old,new}} of the third-order (next-to-next-to-leading
order) QCD corrections to $R_{e^+e^-}$:
\begin{equation}
R_{e^+e^-} \equiv \sigma_{tot}(e^+e^- \rightarrow {\rm hadrons})/
\sigma(e^+e^- \rightarrow \mu^+ \mu^-),
\end{equation}
provides valuable empirical information on the behaviour of perturbation
theory in QCD.  This paper is concerned with ``optimized perturbation
theory'' (OPT) {\cite{OPT}}, and is motivated by three questions which the
$R_{e^+e^-}$ calculation can answer: {\cite{letter}}
\vspace*{2.5mm}

(1)  {\it Does perturbation theory seem to be well behaved?}  Is the
third-order ``optimized'' result in reasonable agreement with the second-order
``optimized'' result?  What can we learn about the error estimate?

(2)  {\it Is the optimized couplant, $a \equiv \alpha_s/\pi$, smaller in
third order than in second?}  The ``induced-convergence'' picture
{\cite{optult}} suggests that the optimized couplant $\bar{a}^{(n)}$, as
determined by the $n$th-order optimization equations, will tend to decrease
as the order $n$ increases.  In this way ``optimization'' could lead to a
convergent sequence of perturbative approximations, even if the
perturbation series in any {\it fixed} renormalization scheme is divergent
{\cite{optult}}.

(3)  {\it Does one find infrared fixed-point behaviour?} A third-order
calculation is a prerequisite for addressing this question in ``optimized''
perturbation theory, and the answer basically depends upon whether the
invariant $\rho_2$ (defined below) is negative or not {\cite{KSS}}.
\vspace*{2.5mm}

  It is striking that, with the originally published third-order result
{\cite{old}}, the answer to all three questions was ``No'' --- while, with
the new, corrected, result {\cite{new}} the answer to all three questions
is ``Yes''.  The purpose of this paper is to elaborate on these three
points, and especially to discuss the infrared fixed-point behaviour
\cite{chyla2,us}.  We do not share the pessimistic attitude of Ch\'yla
{\it et al} \cite{chyla2,higgs} to the infrared results.  If one believes
in OPT, the infrared results -- though quantitatively uncertain -- are
qualitatively unequivocal:  We propose to take them at face value and compare
them to experimental data \cite{us}.

   The plan of this paper is as follows:  Sect.\ 2 reviews OPT, and applies
it to $R_{e^+e^-}$ in third-order, with particular emphasis on the infrared
limit.  Sect.\ 3 compares the predicted $R_{e^+e^-}$ with experimental data,
using Poggio-Quinn-Weinberg (PQW) smearing \cite{PQW}.  Sect.\ 4 briefly
discusses the phenomenology of a ``frozen'' couplant.  Conclusions are
summarized in Sect.\ 5.  Some technical matters are relgated to the appendices.

\section{Optimized Pertubation Theory and Fixed-Point Behaviour}
\setcounter{equation}{0}
\subsection{The principle of minimal sensitivity}

We begin with a few words about the principle of minimal sensitivity, upon
which OPT is based.  It deals with any situation where an exact result is
known to be independent of certain variables, but where the corresponding
approximate result depends upon those variables, and hence is ambiguous.
(In the QCD context, physical quantities are Renormalization Group (RG)
invariant \cite{SP}, but perturbative approximations to them are not, due to
truncation of the perturbation series.)  The philosophy is that such a
non-invariant approximant is most believable where it is least sensitive to
small variations in the extraneous variables, because this is where it best
approximates the exact result's vital property of being completely insensitive
to the extraneous variables.

    A simple example is perhaps the best way to convey this idea.  Consider
the quantum-mechanical problem of computing the eigenvalues, $E_k$, of the
quartic-oscillator Hamiltonian:
\begin{equation}
\label{QHO}
H = {\scriptstyle \frac{1}{2}} p^2 + \lambda x^4,
\end{equation}
where $[x,p] = i$.  Suppose we do standard perturbation theory, but with
\cite{CK}:
\begin{equation}
H_0 = {\scriptstyle \frac{1}{2}}(p^2 + \Omega^2 x^2), \quad\quad
H_{{\rm int}} = \lambda x^4 - {\scriptstyle \frac{1}{2}} \Omega^2 x^2.
\end{equation}
This introduces an ``extraneous variable'' $\Omega$, and the approximate
eigenvalues so calculated will be $\Omega$-dependent.  For example,
first-order for the $k$th eigenvalue gives:
\begin{equation}
\label{E1}
E^{(1)}_k = \frac{1}{2} (k + {\scriptstyle \frac{1}{2}}) \Omega +
\frac{3 \lambda}{4 \Omega^2}(2 k^2 + 2 k + 1).
\end{equation}
However, we {\it know} that the exact eigenvalues are $\Omega$-independent.
Therefore, it is sensible to choose $\Omega$ so that the approximant,
$E^{(1)}_k$, is minimally sensitive to $\Omega$; {\it i.e.},
\begin{equation}
\label{Omega}
\bar{\Omega} = \left[ 3 \lambda \, \frac{(2k^2 + 2k + 1)}
{(k+{\scriptstyle \frac{1}{2}})} \right]^{\frac{1}{3}}.
\end{equation}
(Quite generally, we shall use an overbar to denote an ``optimized'' value.)
This gives the ``optimized'' result:
\begin{equation}
\label{E1opt}
E_k^{(1)}({\rm opt.}) = \frac{3}{4} (k+{\scriptstyle \frac{1}{2}})
\left[ 3 \lambda \, \frac{(2k^2 + 2k + 1)}
{(k+{\scriptstyle \frac{1}{2}})} \right]^{\frac{1}{3}}.
\end{equation}
This simple formula fits the ground-state energy to 2\%, and {\it all} other
energy levels to within 1\%.  The secret of this success is the ``optimal''
choice of $\Omega$, which is different for different levels.

    One may proceed to the calculation of higher-order corrections for some
specific eigenvalue ({\it e.g.}, the ground state, $k\!=\!0$).  For any fixed
$\Omega$ the perturbation series would diverge, but if $\Omega$ is chosen
in each order according to the ``minimal sensitivity'' criterion (which
gives an $\bar{\Omega}$ that gradually increases with order), one finds quite
nice convergence \cite{CK}.  This is an example of the ``induced convergence''
mechanism \cite{optult,indcon}.

    One may also use the same method to obtain accurate approximate
wavefunctions, $\psi_k(x)$, from first-order perturbation theory \cite{KP}.
Here the ``optimal'' $\Omega$ will be a function of $x$; in particular,
it will be proportional to $|x|$ at large $|x|$, thereby converting the
Gaussian dependence $\exp(- \frac{1}{2} \Omega x^2)$ into the correct
large-$|x|$ behaviour.  A variety of other examples and applications can be
found in Refs.\ \cite{OPT,indcon,examples}.  Some examples of QCD applications
can be found in Refs.\ \cite{duke,aurenche}.

\subsection{RG invariance and optimization}

     We next review ``optimized perturbation theory'' (OPT) \cite{OPT} as
applied to the QCD corrections to the $R_{e^+e^-}$ ratio.  Ignoring quark
masses for the present, we may write
$R_{e^+e^-} = 3 \sum q_i^2 (1 + {\cal R})$,
where ${\cal R}$ has the form:
\begin{equation}
{\cal R} = a(1 + r_1 a + r_2 a^2 + \ldots),
\end{equation}
and depends upon a single kinematic variable, $Q$, the {\it cm} energy.  OPT
is based on the fundamental notion of RG invariance \cite{SP},
which means that a physical quantity is independent of the renormalization
scheme (RS).  Symbolically, we can express this by:
\begin{equation}
 0 = \frac{d{\cal R}}{d(RS)} =  \frac{\partial {\cal R}}{\partial (RS)}   +
\frac{da}{d(RS)} \frac{\partial {\cal R}}{\partial a}, \label{rg}
\end{equation}
where the total derivative is separated into two pieces corresponding to
RS dependence from the series coefficients, $r_i$, and from the couplant,
respectively.  A particular case of Eq. (\ref{rg}) is the familiar equation
expressing the renormalization-scale independence of ${\cal R}$:
\begin{equation}
\left( \mu \frac{\partial}{\partial \mu} +
\beta(a) \frac{\partial}{\partial \mu} \right) {\cal R} = 0,
\end{equation}
where
\begin{equation}
\beta(a) \equiv \mu \frac{da}{d\mu} = - ba^2(1+ca + c_2a^2+\ldots).
\end{equation}
The first two coefficients of the $\beta$ function are RS invariant
and, in QCD with $N_f$ massless flavours, are given by:
\begin{equation}
b = \frac{(33-2N_f)}{6}, \quad\quad c= \frac{153 - 19 N_f}{2(33-2N_f)}.
\end{equation}

     When integrated, the $\beta$-function equation can be written as:
\begin{equation}
\label{beta}
\int_0^{a} \frac{d a^{\prime}}{ \beta( a^{\prime})} + {\cal C} =
\int_{\tilde{\Lambda}}^{\mu}
\frac{d \mu^{\prime}}{\mu^{\prime}} = \ln(\mu/{\tilde{\Lambda}}).
\end{equation}
where ${\cal C}$ is a suitably infinite constant and $\tilde{\Lambda}$ is a
constant with dimensions of mass.  The particular definition of
$\tilde{\Lambda}$ that we use corresponds to choosing \cite{OPT}
\begin{equation}
{\cal C} = \int_0^{\infty} \frac{d a^{\prime}}
{ b {a^{\prime}}^2(1 + c a^{\prime})}
\end{equation}
(where it to be understood that the integrands on the left of (\ref{beta})
are to be combined before the bottom limit is taken).  This
$\tilde{\Lambda}$ parameter is related to the traditional definition
\cite{BBDM} by an RS-invariant, but $N_f$-dependent factor:
\begin{equation}
\ln (\Lambda/\tilde{\Lambda}) = (c/b) \ln (2c/b).
\end{equation}
The $\Lambda$ parameter is scheme dependent, but the $\Lambda$'s of different
schemes can be related exactly by a 1-loop calculation \cite{CG,fncg}.  As is
usual, we shall regard $\Lambda_{\overline{\rm MS}}$ (for 4 flavours) as the
free parameter of QCD \cite{fnlam}.

    From Eq. (\ref{beta}) it is clear that $a$ depends on RS only through
the variables $\mu/\tilde{\Lambda}$ and $c_2, c_3, \ldots$, the
scheme-dependent $\beta$-function coefficients.  The coefficients of
${\cal R}$ can depend on RS only through these same variables, because of
RG invariance, Eq. (\ref{rg}).  Therefore, these variables provide a complete
RS parametrization, as far as physical quantities are concerned \cite{OPT}.
Thus, we may write:
\begin{equation}
 a=a({\rm RS})=a(\tau,c_2,c_3,\ldots),
\end{equation}
where
\begin{equation}
\tau \equiv b \ln(\mu/\tilde{\Lambda}).
\end{equation}
The $\tau$ variable is convenient and also serves to emphasize the very
important point that RS dependence involves only the {\it ratio} of $\mu$
to $\tilde{\Lambda}$.  ``Optimization'' does {\it not} determine an
``optimum $\mu$'', but it will determine an optimum $\tau$.

   The dependence of $a$ on the set of RS parameters $\tau$ and $ c_j$
\cite{OPT} is most easily obtained \cite{pol} by taking partial derivatives
of Eq. (\ref{beta}), varying one parameter while holding the others constant.
This yields:
\begin{equation}
\frac{\partial a}{\partial \tau} = \beta(a)/b,
\end{equation}
\begin{equation}
\label{betaj}
\frac{\partial a}{\partial c_j} \equiv \beta_j(a)=
-b \beta(a) \int_0^a dx \; \frac{x^{j+2}}{[\beta(x)]^2}.
\end{equation}
Note that the $\beta_j$ functions begin at order $a^{j+1}$.

    The symbolic RG-invariance equation (\ref{rg}) can now be written out
explicitly as the following set of equations:
\begin{eqnarray}
\label{rga}
\left( \left. \frac{\partial}{\partial \tau} \right|_a +
\frac{\beta(a)}{b} \frac{\partial}{\partial a} \right) {\cal R} \, = 0,  & & \\
\label{rgb}
\left( \left. \frac{\partial}{\partial c_j} \right|_a +
\beta_j(a) \frac{\partial}{\partial a} \right) {\cal R} = 0 & \quad\quad
{\scriptstyle (j = 2,3,\ldots)}. &
\end{eqnarray}
These equations determine how the coefficients $r_i$ of ${\cal R}$ must
depend on the RS variables.  Thus, $r_1$ depends on $\tau$ only, while
$r_2$ depends on $\tau$ and $c_2$ only, etc., with
\begin{equation}
\frac{\partial r_1}{\partial \tau} = 1,
\end{equation}
\begin{equation}
\frac{\partial r_2}{\partial \tau} = 2 r_1 + c,  \quad\quad
\frac{\partial r_2}{\partial c_2} = - 1,
\end{equation}
etc..  Upon integration one will obtain
$r_i = f(\tau,c_2,\ldots,c_i) + \mbox{const.}$, where $f$ is a known function
and the constant of integration is an RS invariant.  Thus, certain combinations
of series coefficients and RS parameters:
\begin{equation}
\label{rho1}
\rho_1(Q) \equiv \tau - r_1,
\end{equation}
\begin{equation}
\label{rho2}
\rho_2 \equiv r_2 + c_2 -(r_1+\frac{1}{2}c)^2,
\end{equation}
etc., are RS invariant \cite{OPT,fnterho2}.  In the $e^+e^-$ case,
$\rho_1$ is a function of the {\it cm} energy $Q$, while $\rho_2$ and
the higher-order invariants are pure numbers, dependent only on the number
of flavours, $N_f$.

    Although the exact ${\cal R}$ is RG-invariant, the truncation of the
perturbation series spoils this invariance.  The $n$th-order approximant
${\cal R}^{(n)}$, defined by truncating ${\cal R}$ and the $\beta$ function
to only $n$ terms, depends on the RS variables $\tau,\ldots,c_{n-1}$.
OPT corresponds to choosing an ``optimal'' RS in which the approximant
${\cal R}^{(n)}$ is stationary with respect to RS variations; {\it i.e.},
the RS in which ${\cal R}^{(n)}$ {\it exactly} satisfies the RG-invariance
equations, (\ref{rga}, \ref{rgb}).  (Note that only the first $(n-1)$
equations will be nontrivial in $n$th order.)

   The second order approximant is
\begin{equation}
{\cal R}^{(2)} = a(1 + r_1 a),
\end{equation}
where $a$ here is short for $a^{(2)}$, the solution to (\ref{beta}) with
$\beta$ truncated at second order.  ${\cal R}^{(2)}$ depends on RS only
through the variable $\tau$.  The optimization equation, from (\ref{rga}), is
\begin{equation}
\bar{a}^2 - \bar{a}^2(1 + c \bar{a})(1 + 2 \bar{r}_1 \bar{a}) = 0.
\end{equation}
This equation, together with the $\rho_1$ definition and the second-order
integrated $\beta$-function equation, (\ref{beta}), uniquely determines the
optimized result.  (For details, see \cite{OPT,duke}.)

   The third order approximant is:
\begin{equation}
{\cal R}^{(3)} = a(1 + r_1 a + r_2 a^2),
\end{equation}
where now $a$ is short for $a^{(3)}$, the solution to (\ref{beta}) with
$\beta$ truncated at third order.  ${\cal R}^{(3)}$ depends on RS through two
parameters $\tau$ and $c_2$, so there are two optimization equations, coming
from (\ref{rga}, \ref{rgb}).  These can be reduced to \cite{OPT}:
\begin{equation}
\label{eq1}
(3\bar{r}_2 + 2\bar{r}_1c + \bar{c}_2) +
(3\bar{r}_2 c + 2 \bar{r}_1 \bar{c}_2) \bar{a} +
3 \bar{r}_2 \bar{c}_2 \bar{a}^2 = 0,
\end{equation}
\begin{equation}
\label{eq2}
I(1+(c+2\bar{r}_1)\bar{a}) - \bar{a} = 0, \;\;\;\;\;\;
\mbox{where} \;\;\;\;
I= \int_0^{\bar{a}} \frac{dx}{(1+cx + \bar{c}_2 x^2)^2}.
\end{equation}
This integral can be done analytically, and is given by:
\begin{equation}
 I = \frac{1}{\Delta^2} \left[
\frac{\bar{a}(c^2 -2 \bar{c}_2 + c \bar{c}_2 \bar{a})}
{(1+c \bar{a} + \bar{c}_2 \bar{a}^2)}
- 4 \bar{c}_2 f(\bar{a},\bar{c}_2) \right],
\end{equation}
with
\begin{equation}
\label{f}
f(\bar{a},\bar{c}_2) = \frac{1}{2\Delta}
\ln \left[ \frac{1+ {\scriptstyle \frac{1}{2}} \bar{a}(c+\Delta)}
{1+ {\scriptstyle \frac{1}{2}} \bar{a}(c-\Delta)} \right],
\end{equation}
where $\Delta^2 \equiv c^2-4\bar{c}_2$.  (This assumes that $\Delta^2 > 0$,
which proves to be true here.)  The procedure for solving these optimization
equations is discussed further in subsection 2.4, but we next discuss the
infrared limit.

\subsection{Infrared limit and fixed-point behaviour}

      Suppose we consider ${\cal R}$ at lower and lower {\it cm} energy, $Q$.
Since $c$ is positive for $N_f \! \le \! 8$, the second-order $\beta$
function has no non-trivial zero.  Thus, in any RS, the couplant $a^{(2)}$
and approximant ${\cal R}^{(2)}$ must become singular at some $Q$ of order
$\Lambda_{\overline{\rm MS}}$.  In third order this may or may not happen,
depending on whether the RS has a positive or negative $c_2$.  If $c_2$ is
negative then the couplant remains finite and tends to a ``fixed-point''
value, $a^*$, which is the non-trivial zero of the third-order $\beta$
function; {\it i.e.,} the positive root of
\begin{equation}
\label{betaz}
1 + c a^* + c_2 a^{*2} = 0.
\end{equation}
Since fixed-point behaviour hinges on $c_2$, which is scheme dependent,
it is vital to have a sensible choice of RS \cite{KSS}.  In OPT the optimal
$c_2$ is determined by the optimization equations, and depends on $Q$
somewhat.  If $\bar{c}_2$ is negative as $Q \to 0$, then OPT will give
fixed-point behaviour.  The infrared limit of the optimization process
was analyzed in Ref.\ \cite{KSS} and we briefly review the relevant results.

  Since $\beta$ vanishes at a fixed point, the $\tau$ optimization equation,
corresponding to (\ref{rga}), reduces to
\begin{equation}
1 + (2 \bar{r}_1 + c)\bar{a}^* = 0.
\end{equation}
Then, just by differentiating (\ref{betaz}) with respect to $c_2$, one
obtains:
\begin{equation}
\lim_{a \rightarrow a^*} \beta_2^{(3)}(a) =
\frac{\partial a^*}{\partial c_2}
= \frac{- a^{*2}}{(c + 2 c_2 a^*)}.
\end{equation}
(This can also be obtained, more laboriously, as the limit of (\ref{betaj}).)
Thus, the $c_2$ optimization equation, corresponding to (\ref{rgb}), becomes
\begin{equation}
\bar{a}^* + \frac{(1+ 2 \bar{r}_1 \bar{a}^* + 3 \bar{r}_2 \bar{a}^{*2})}
{(c+2 \bar{c}_2 \bar{a}^*)} =0.
\end{equation}
The two optimization equations yield:
\begin{equation}
\label{fpr}
\bar{r}_1 = - \frac{1}{2} \frac{(1+c \bar{a}^*)}{\bar{a}^*}, \quad\quad
\bar{r}_2 = - \frac{2}{3} \bar{c}_2.
\end{equation}
Using the expression for the invariant $\rho_2$, (\ref{rho2}), one obtains:
\begin{equation}
\label{fpeqc2}
\bar{c}_2 = 3(\rho_2 + \frac{1}{4 \bar{a}^{*2}}).
\end{equation}
Finally, substituting into the fixed-point condition (\ref{betaz}), one finds
\cite{KSS}:
\begin{equation}
\label{fpeq}
\frac{7}{4} + c \bar{a}^* + 3 \rho_2 \bar{a}^{*2} = 0,
\end{equation}
which determines $\bar{a}^*$ in terms of the RS-invariant quantities $c$ and
$\rho_2$.  A positive $\bar{a}^*$ exists if $\rho_2$ is negative, and the
more negative $\rho_2$ is, the smaller $\bar{a}^*$ will be.

\subsection{Implementing the optimization procedure}

    Returning to finite $Q$, we now consider how to obtain the third-order
optimized approximant $\bar{{\cal R}}^{(3)}$ numerically as a function of $Q$.
As input, we need the values of $\rho_1$ and $\rho_2$.  Being invariants,
they can be obtained from calculations performed in any computationally
convenient RS.  The calculations in the literature have used the ``modified
minimal subtraction'' ($\overline{\rm MS}$) convention, with the
renormalization point $\mu$ chosen to be $Q$.  The ${\cal R}$ coefficients
are \cite{r1,new}:
\begin{equation}
r_1({\scriptstyle \overline{\rm MS}; \; \mu = Q}) =
1.9857 - 0.1153 N_f,
\end{equation}
\begin{equation}
r_2({\scriptstyle \overline{\rm MS}; \; \mu = Q}) =
-6.6368 - 1.2001 N_f - 0.0052 N_f^2
- 1.2395 \: {( {\textstyle \sum} q_i )}^2 \! / (3 {\textstyle \sum} q_i^2 ).
\end{equation}
The RS parameters of the $\overline{\rm MS}(\mu \! =\! Q)$ scheme are:
\begin{equation}
\tau({\scriptstyle \overline{\rm MS}; \; \mu = Q})
= b \ln (Q/\tilde{\Lambda}_{\overline{\rm MS}})
= b \ln (Q/\Lambda_{\overline{\rm MS}}) + c \ln (2c/b),
\end{equation}
\begin{equation}
c_2({\scriptstyle \overline{\rm MS}}) = \frac{3}{16} \frac{1}{(33-2N_f)}
\left[ \frac{2857}{2} - \frac{5033}{18} N_f + \frac{325}{54} N_f^2 \right].
\end{equation}
(The latter was first calculated in Ref.\ \cite{c2}, and has recently been
confirmed independently \cite{c2new}.)  Substitution of these results
into (\ref{rho1}, \ref{rho2}) gives the invariants.  One can see explicitly
that $\rho_1$ depends logarithmically on the {\it cm} energy $Q$, and on the
free parameter of QCD, $\Lambda_{\overline{\rm MS}}$ \cite{fnlam}.  However,
$\rho_2$ depends only on $N_f$.  Since $\rho_2$ turns out to be negative, one
will find fixed-point behaviour in the ``optimum'' scheme \cite{fac,fna}.
Table 1 gives the fixed-point couplant values, determined from Eq.
(\ref{fpeq}), for various $N_f$ values \cite{fnb}.

     Consider a world with $N_f$ massless quarks, ignoring complications
due to quark thresholds for the present.  For simplicity we assume that the
value of $\Lambda_{\overline{\rm MS}}$ is given, and our numerical results
use $\Lambda_{\overline{\rm MS}} = 230$ MeV for four flavours \cite{fnrep}.
For any chosen $Q$ we then have definite numerical values for the invariant
quantities $\rho_1$, $\rho_2$ (and $b$, $c$).  We need to solve for the
optimum couplant, $\bar{a}$, and the optimized coefficents, $\bar{r}_1$,
$\bar{r}_2$, and this will involve determining the RS parameters $\bar{\tau}$,
$\bar{c}_2$ of the optimal RS.  These five variables are related by five
equations; the two optimization equations, (\ref{eq1}, \ref{eq2}), the
$\rho_1$, $\rho_2$ equations, (\ref{rho1}, \ref{rho2}), and the integrated
$\beta$-function equation, (\ref{beta}), which for the $\beta$-function
truncated at third order becomes explicitly:
\begin{equation}
\label{k3}
\tau = \frac{1}{a} + c \ln (ca) - \frac{1}{2} c \ln (1+ca+c_2 a^2)
- (c^2 - 2 c_2) f(a,c_2),
\end{equation}
where $f(a,c_2)$ is given by (\ref{f}).  By substituting this last equation
into the $\rho_1$ equation we can obtain $\bar{r}_1$ explicitly as a function
of $\bar{a}$, $\bar{c}_2$.  We can then rearrange the $\rho_2$ equation
to give $\bar{r}_2$ explicitly as a function of $\bar{a}$, $\bar{c}_2$.
This leaves $\bar{a}$, $\bar{c}_2$ to be solved for from the two optimization
equations.  Starting from an initial guess for $a$ and $c_2$, our procedure
was to solve (\ref{eq2}) numerically for a new $\bar{a}$; and then (with the
new $\bar{a}$) to solve (\ref{eq1}) for a new $\bar{c}_2$.  We then iterated
this procedure until the difference between successive solutions reached a
specified tolerance.  Further details are given in Appendix A.

At very low $Q$ we encountered technical problems with slow convergence
of the iteration scheme.  These are discussed in Appendix A.  Nevertheless,
with care it was possible to obtain accurate solutions at low energies.
In Fig.\ 1 we show the optimized solution in the $a$, $c_2$ plane
as it smoothly approaches the fixed-point solution, which lies on the infrared
boundary $1+ca+c_2 a^2 = 0$.  The figure shows two cases, $N_f=3$ and $N_f=2$.
(In the real-world case we must switch from 3 flavours to 2 when we cross the
strange-quark threshold.  This requires a matching of $\Lambda$ parameters,
as discussed in Appendix B.)

   The optimized couplant $\bar{a}$ is shown as a function of $Q$ in
Fig.\ 2.  Note that the effective couplant below 300 MeV is nearly constant
at about 0.263, which is the $N_f=2$ fixed-point value.  Fig.\ 2 also shows
the second- and third-order optimized results for ${\cal R}$.  The
second-order result diverges at $Q \approx 400$ MeV, where $\rho_1(Q)$
vanishes.  However, $\bar{{\cal R}}^{(3)}$ remains finite, rising only to
0.33 at $Q=0$.

\subsection{Illustrative results}

   We pause for a moment to consider a comparison between the second- and
third-order optimized results at moderately high $Q$.  This exercise was
performed by several authors \cite{max,oldopt} when the `old' third-order
result \cite{old} was first published, and the results were disquieting.
However, the new result \cite{new} has transformed the situation, which is
now very satisfactory.  In Table 2 we give details for the two illustrative
cases considered by Maxwell and Nicholls \cite{max}, namely $N_f = 5$,
$Q = 34$ GeV, with either ${\tilde{\Lambda}}_{\overline{\rm MS}} = 100$ MeV
or ${\tilde{\Lambda}}_{\overline{\rm MS}} = 500$ MeV.
[Note, though, that the results depend only on the ratio of $Q$ to
${\tilde{\Lambda}}_{\overline{\rm MS}}$.]  From Table 2 we see that between
second and third order the optimized prediction $\bar{{\cal R}}$ decreases
only a few percent.  With the `old' result there had appeared to be a
disconcertingly large increase \cite{max,oldopt}.  The new situation is much
more satisfactory in other ways, too:  In both examples the coefficient
$\bar{r}_2$ now has a more reasonable magnitude, and the $\bar{r}_1$
coefficient has not changed so drastically from second to third order.

   The optimized couplant $\bar{a}$ now shows a marked decrease from second
to third order.  This is just what one would expect in the ``induced
convergence'' picture of Ref.\ \cite{optult}.  In that picture ``optimization''
induces convergence through a mechanism in which the effective expansion
parameter, $\bar{a}$, shrinks from one order to the next.  Note that the
`old' result gave the opposite behaviour, with $\bar{a}$ apparently
increasing from second to third order.

   The results also shed some light on the error-estimation question:  If we
knew just $\bar{{\cal R}}^{(2)} = \bar{a}(1 + \bar{r}_1 \bar{a})$, how might
we estimate the error?  Two estimates suggest themselves: (i) $n \bar{a}^3$,
where $n$ is an order-one number, which presumes a well-behaved converging
series, or (ii) $| \bar{r}_1 \bar{a}^2 |$, the magnitude of the last
calculated term, which is a typical error estimate for an asymptotic series.
Knowing $\bar{{\cal R}}^{(3)}$ we can, presumably, get a much better idea of
the actual error in $\bar{{\cal R}}^{(2)}$ from the difference
$\delta \equiv \bar{{\cal R}}^{(3)} - \bar{{\cal R}}^{(2)}$.
We have compared $\delta$ with estimates (i) and (ii) over a wide range of
$Q/\Lambda_{\overline{\rm MS}}$ values.

   Estimate (i), if we had assumed $n \approx 1$ or 2, would
have been rather too optimistic.  In fact, $| \delta |$ is between 7 and 14
times $\bar{a}^3$ (for $N_f = 4$ and $Q/\Lambda_{\overline{\rm MS}} \ge 5$).
This is directly related to the size of the invariant $\rho_2$ (which is about
$-14$ for 4 flavours).  [One can show analytically that
$\delta = \rho_2 \bar{a}^3 + {\cal O}(\bar{a}^4)$ in the large-$Q$ limit.]
Of course, we could not know $\rho_2$ until a third-order calculation was
done.  Arguably, though, 14 can still be considered an ``order-one'' number,
especially in a theory that naturally involves numbers such as 4 (flavours),
3 (colours), 8 (gluons), etc..

   Estimate (ii), based on the last calculated term, agrees with $| \delta |$
to within a factor of two either way for $Q/\Lambda_{\overline{\rm MS}}$
between 5 and 1000.  At higher $Q/\Lambda_{\overline{\rm MS}}$ values this
estimate would be overly pessimistic.  However, we think that for present
energies the estimate (ii) is perhaps the safest way to estimate the error.
We suggest that it be used in QCD applications where only second-order results
are known.

\subsection{Credibility of the infrared results}

   We have stressed that OPT yields finite results for ${\cal R}$ down to
$Q=0$.  The crucial question is, of course: How meaningful are these results?
We would like to explain why, in contrast to other authors
\cite{chyla2,higgs,chyla}, we take a positive attitude on this issue.

   Firstly, suppose we adopt the philosophy that the last calculated term
in the ``optimized'' perturbation series is a measure of the error.
As we saw in this last section, this proved to be reasonable in the
second-order case.  In third order this gives $|\bar{r}_2 \bar{a}^3|$ as
the error estimate.  Since $\bar{{\cal R}} \approx \bar{a}$, this implies
a fractional error of $|\bar{r}_2 \bar{a}^2|$.  In Table 3 we show some
illustrative results at low energies, together with their estimated error.
{}From this one can see that the behaviour of the series is quite satisfactory
above $Q=1$ GeV.  The situation undoubtedly deteriorates at lower energies; by
the time we reach $Q=0$ we have a series of the form $0.26(1-0.76+1.01)$ in
which the higher-order terms are comparable to the leading term.  While this
is hardly a good situation, it is not completely disastrous; the corrections
alternate in sign, and they do not dwarf the leading term.  We believe that
our error estimate, which grows to 100\% at $Q=0$, is not unreasonable: the
result may well be off by a factor of 2, but is unlikely to be off by an order
of magnitude.  The qualitative conclusion, that ${\cal R}$ remains small
(say, $0.3 \pm 0.3$) at low energies is hard to escape.

    Secondly, it is instructive to view the use of QCD perturbation theory
in the infrared limit as an extrapolation away from $N_f = 33/2$ \cite{banks}.
At $N_f = 33/2$ the leading $\beta$-function coefficient, $b$, vanishes (and
hence $c$ goes to $- \infty$).  For $N_f = 33/2 - \epsilon$, with $\epsilon$
small and positive, there must be an infrared fixed point at
$a^* \sim - 1/c = {\cal O}(\epsilon)$ \cite{banks}.  Perturbative
calculations, even in the infrared, should then be meaningful if $\epsilon$
is sufficiently small.  Furthermore, one could naturally expect that the
more orders in perturbation theory one has, the further one can extrapolate
from $N_f = 33/2$.  With sufficient orders one should be able to get
infrared results down to $N_f =0$, unless there is some unknown reason for
the behaviour of the theory to change fundamentally at some critical $N_f$
between $33/2$ and 0.  What does happen?  Well, at second order, of course,
one finds fixed-point behaviour, with $a^* = -1/c$, provided $c$ is negative,
which requires $N_f > 153/19 \approx 8$, though $N_f$ needs to be still larger
if $a^*$ is to be reasonably small.  In third order our results imply that,
in the $R_{e^+e^-}$ case, fixed-point behaviour --- with moderately small
$a^*$ values --- does extend to $N_f=0$.

    In the $\epsilon \to 0$ limit, $a^*$ tends to $-1/c$, and hence to
$(8/321) \epsilon$.  The small coefficient suggests that the natural expansion
parameter of an extrapolation from $N_f = 33/2$ is not $\epsilon$ but
approximately $\epsilon/40$.  One can verify that the third-order OPT results
smoothly approach the limiting form as $\epsilon \to 0$.  For $N_f = 16$ one
has
$\rho_2 = -1724.4$, and one gets a series of the form $0.012(1 - 0.03 + 0.04)$.
As $N_f$ decreases, the behaviour of the series deteriorates, but it does so
quite steadily;  there is no dramatic change around $N_f =8$ or any other
$N_f$.

   In conclusion, our view is that the $N_f = 2$ infrared results, while
quantitatively uncertain, are qualitatively credible.  Having made this case
in theoretical terms, let us now see what experiment has to say.

\section{Comparing Theory to Experiment}

\subsection{$R_{e^+e^-}$ including quark masses}

     In this section we construct the theoretical prediction for $R_{e^+e^-}$
(allowing for quark masses) and discuss its comparison with experiment using
the PQW smearing method.  We limit ourselves to the region below 6 GeV, and
we shall be particularly interested in the region below 1 GeV.

  To allow for quark masses in $R_{e^+e^-}$, we used the following approximate
formula \cite{PQW}:
\begin{equation}
R_{e^+e^-} = 3 \sum_i q_i^2 \; T(v_i)[ 1 + g(v_i) {\cal R}],
\end{equation}
where the sum is over all quark flavours that are above threshold ({\it i.e.},
whose masses, $m_i$, are less than $Q/2$), and
\begin{eqnarray}
v_i  & = & (1-4 m_i^2/Q^2)^{\frac{1}{2}}, \nonumber \\
T(v) & = & v(3-v^2)/2, \\
g(v) & = & \frac{4 \pi}{3} \left[ \frac{\pi}{2v} -
\frac{(3+v)}{4} \left( \frac{\pi}{2} - \frac{3}{4 \pi} \right) \right].
\nonumber
\end{eqnarray}
The coefficient $T(v_i)$ is the parton-model mass dependence and $g(v_i)$
is a convenient approximate form for the mass dependence of the leading-order
QCD correction \cite{PQW,schwinger}.  The higher-order corrections have been
calculated only for massless quarks, so we simply evaluate ${\cal R}$ with
$N_f$ equal to the the number of above-threshold flavours.

   In our numerical results we used standard values for the current-quark
masses \cite{PDG}: $m_u = 5.6$ MeV, $m_d = 9.9$ MeV, $m_s = 199$ MeV,
$m_c = 1.35$ GeV.  For $\Lambda_{\overline{\rm MS}}$ we used a 4-flavour
value of 230 MeV above charm threshold ($Q > 2 m_c$).  Then, each time a
flavour threshold was crossed as we decreased $Q$, we reduced $N_f$ by 1 and
computed the new $\Lambda_{\overline{\rm MS}}$ parameter appropriate to the
new $N_f$.  The matching of $\Lambda$'s is discussed in Appendix B.

    In this way we obtained the ``raw'' theoretical prediction for
$R_{e^+e^-}$ shown in Fig.\ 3.  For comparison, the figure also shows the
parton-model result ({\it i.e.,} with the QCD correction term ${\cal R}$
set to zero).

\subsection{PQW smearing}

   A {\it direct} comparison of the theoretical prediction with the
experimental data is not possible, because there is no direct correspondence
between the perturbative quark-antiquark thresholds and the hadronic
thresholds and resonances of the data.  However, a meaningful comparison is
possible if some kind of ``smearing'' procedure is used \cite{PQW,barnett}.
We used the smearing method of Poggio, Quinn, and Weinberg (PQW) \cite{PQW},
who define the ``smeared'' quantity:
\begin{equation}
\label{PQWeq}
\bar{R}_{\scriptscriptstyle{PQW}}(Q;\Delta) =
\frac{\Delta}{\pi} \int_0^{\infty} ds^{\prime}
\frac{R_{e^+e^-}(\sqrt{s^{\prime}})} {(s^{\prime} - Q^2)^2 + \Delta^2}.
\end{equation}
In terms of the vacuum-polarization amplitude $\Pi$, one can write
$\bar{R}_{\scriptscriptstyle{PQW}}$ as
\cite{PQW}:
\begin{equation}
2 i \bar{R}_{\scriptscriptstyle{PQW}}(Q;\Delta) =
\Pi(Q^2 + i \Delta) - \Pi(Q^2 - i \Delta).
\end{equation}
In the limit $\Delta \to 0$ this reduces to $2 i R_{e^+e^-}$, which is the
discontinuity of $\Pi$ across its cut.  However, a finite $\Delta$ keeps one
away from the infrared singularities and nonperturbative effects that lurk
close to the cut.  The idea is to apply this smearing to both the theoretical
and experimental $R_{e^+e^-}$'s and then compare them.

In principle, the more orders in perturbation theory one has, the smaller
one can take $\Delta$ \cite{PQW}.  However, this requires the full mass
dependence of the higher-order corrections, which we do not know.  In their
leading-order study of the charm-threshold region, PQW used a value
$\Delta = 3$ GeV$^2$, and we shall use values of the same order of magnitude.
We take a pragmatic view: the best choice of $\Delta$ is the smallest value
that will smooth out any rapid variations in either the experimental or
the theoretical $R_{e^+e^-}$.  It turns out that this depends upon the energy
region one is interested in.  Around charm threshold a $\Delta$ of 3 GeV$^2$
or more is necessary, while in the lowest energy region a $\Delta$ as small
as 1 GeV$^2$ can be used.

  The integral in Eq. (\ref{PQWeq}) was evaluated by numerical integration,
after first making a change of variables $s^{\prime} - Q^2 = \Delta
\tan\theta$.
The computer routine was designed to take an input $R_{e^+e^-}$, specified
over a range 0 to $Q_{{\rm max}}$ and to evaluate the integral over this
range.  A term was then added to account for the contribution from
$Q_{{\rm max}}$ to $\infty$, assuming that $R_{e^+e^-}$ remained constant above
$Q_{{\rm max}}$.  The accuracy of the numerical-integration routine was tested
against analytic results for several simple input functions.

\subsection{Experimental data and resonances}

    The experimental data we used comes from a variety of sources:
$e^+e^- \to \pi^+\pi^-$ data in the $\rho$ region and above from the
OLYA/CMD and DM2 collaborations \cite{barkov,dm2}; Adone $\gamma\gamma 2$
data from 1.4 to 3 GeV \cite{bacci}; SLAC Mark I data from 3 to 6 GeV
\cite{siegrist}; and Crystal Ball data above 5 GeV \cite{crystal}.
For useful compilations and reviews see Ref.\ \cite{reviews}.  We used simple
fits to the data in some regions, particularly when the data had a lot of
structure and/or had large statistical errors.  This was more convenient for
the numerical integration routine and made it easier for us to examine the
effect of the experimental uncertainties on the smeared result.  Fig.\ 4 shows
our data compilation, up to 6 GeV, excluding narrow resonances.  In fact, we
used data going well beyond $b$ threshold, but they have no real effect on the
results we present.

    The sharp resonances $\omega$, $\phi$, $J/\psi$, $\psi^{\prime}$, and
$\psi(3770)$ were not included in the data compilation so that their
contribution to $\bar{R}_{\scriptscriptstyle{PQW}}$ could be put in
analytically.  They have a relativistic Breit-Wigner form \cite{collider}:
\begin{equation}
 R_{\scriptscriptstyle{res}} = \frac{9}{\alpha^2} B_{\it ll} B_h
\frac{M^2 \Gamma^2}{(s-M^2)^2 + M^2\Gamma^2}, \label{eq:rres}
\end{equation}
where $M$, $\Gamma$, $B_{\it ll}$, and $B_h$ are, respectively, the mass,
width, leptonic branching fraction, and hadronic branching fraction of the
resonance.  The parameters for the resonances were taken from the 1992 Review
of Particle Properties \cite{PDG}.  For $B_{\it ll}$ we used the weighted
average of the $ee$ and $\mu\mu$ branching ratios.

  The contribution of such a Breit-Wigner resonance to the smearing integral
(\ref{PQWeq}) can be evaluated analytically using partial fractions.  [The
resulting expression is too cumbersome to quote, but we may note that the
narrow width approximation:
\begin{equation}
 \frac{1}{(s-M^2)^2 + M^2\Gamma^2} \approx \frac{\pi}{M\Gamma} \delta(s-M^2),
\end{equation}
which gives a contribution to $\bar{R}_{\scriptscriptstyle{PQW}}$ of
\begin{equation}
\bar{R}_{\scriptscriptstyle{res}} \approx
\frac{9 B_{\it ll} B_h \Delta M \Gamma}{\alpha^2[(s-M^2)^2 + \Delta^2]},
\end{equation}
is a pretty good approximation.]  In Fig.\ 5 we show, for two different
$\Delta$ values, the contributions of the various resonances to the
experimental $\bar{R}_{\scriptscriptstyle{PQW}}$.  The $\rho$'s contribution
is shown by a dotted line.  However, since the $\rho$ is rather wide and
asymmetric, it was actually treated, not in this manner, but by numerical
integration, using the data points from Ref.\ \cite{barkov} as part of the
data compilation (Fig.\ 4).

\subsection{Results and uncertainty estimates}

    The results obtained by applying PQW smearing to both theory and
experiment are shown in Fig.\ 6.  For the smaller $\Delta$ (1 GeV$^2$)
there is good agreement between theory and experiment below 1 GeV, but in the
charm-threshold region there is clearly insufficient smearing for the
comparison to be meaningful.  Increasing $\Delta$ to 3 GeV$^2$ smooths out
the experimental curve almost completely.  The agreement between theory and
experiment is excellent below 2 GeV.  In the charm region the agreement is
less good, but this can be attributed mainly to the sizable systematic
normalization uncertainty (10 -- 20\%) in the data in this region, which
produces an uncertainty of about $\pm 0.4$ in the experimental
$\bar{R}_{\scriptscriptstyle{PQW}}$ at around $Q = 4$ GeV.
For comparison, Fig.\ 6(b) also includes the naive parton-model prediction.
One can see from this that the QCD correction term ${\cal R}$ provides about
a 20\% increase which is vital to the good agreement with the data.

  Using $Q^2$, rather than $Q$ as the variable, we can continue
$\bar{R}_{\scriptscriptstyle{PQW}}(Q^2)$ into the negative $Q^2$ region
(Cf. Ref.\ \cite{adler}).  As shown in Fig.\ 7, for $\Delta = 1$ GeV$^2$,
the good agreement persists.

     To quantify the good agreement at low energies, we discuss how various
uncertainties would affect $\bar{R}_{\scriptscriptstyle{PQW}}$ at $Q=0$.
First we discuss the experimental uncertainties.  There is about a 5\%
uncertainty in the $\rho$, $\omega$ and $\phi$ contributions, due to the
uncertainty in their total and leptonic widths.  For $\Delta=1$ (3) GeV$^2$
this gives an error in $\bar{R}_{\scriptscriptstyle{PQW}}(0)$ of about
$\pm 0.04$ ($\pm 0.02$).  Uncertainties in the $\psi$ resonance parameters
affect $\bar{R}_{\scriptscriptstyle{PQW}}(0)$ by $\pm 0.01$ or less.
We considered the effect of a 15\% normalization change in the continuum
data in the 1.5 -- 3 GeV region:  The effect on
$\bar{R}_{\scriptscriptstyle{PQW}}(0)$ was about $\pm 0.03$ ($\pm 0.06$)
for $\Delta=1$ (3) GeV$^2$.  We also allowed for a 15\% normalization change
in the 3 -- 5 GeV region.  The effect on $\bar{R}_{\scriptscriptstyle{PQW}}(0)$
was about $\pm 0.02$ ($\pm 0.065$) for $\Delta=1$ (3) GeV$^2$.  Combining
these four distinct sources of error in quadrature, we estimate an overall
uncertainty in the experimentally determined
$\bar{R}_{\scriptscriptstyle{PQW}}(0)$ of $\pm 0.06$ for $\Delta=1$ GeV$^2$
and $\pm 0.08$ for $\Delta=3$ GeV$^2$.

   On the theoretical side, errors arise from two sources: (i) uncertainty
in the input parameters (quark masses and $\Lambda_{\overline{\rm MS}}$), and
(ii) truncation of the perturbation series.  We varied each quark mass by its
quoted error \cite{PDG}.  Varying the $u$ and $d$ masses had negligible
effect.  $\bar{R}_{\scriptscriptstyle{PQW}}(0)$ changed by $\pm 0.004$
($\pm 0.001$) on varying the $s$ mass, and by $\pm 0.005$ ($\pm 0.013$) on
varying the $c$ mass for $\Delta=1$ (3) GeV$^2$.  Changing
$\Lambda_{\overline{\rm MS}}$ by 50 MeV to 280 MeV increased
$\bar{R}_{\scriptscriptstyle{PQW}}(0)$ by 0.019 (0.014) for $\Delta=1$ (3)
GeV$^2$.  The series-truncation error can reasonably be estimated from the
last term in the optimized series, as we argued earlier.  At 1 GeV this
suggests that ${\cal R}$ is accurate to about 10\%, and is considerably
more accurate at larger energies.  This is corroborated by the good agreement
between second- and third-order results.  The theoretical uncertainty in
${\cal R}$ above 1 GeV contributes an error in
$\bar{R}_{\scriptscriptstyle{PQW}}(0)$ of less than $\pm 0.006$ ($\pm 0.009$)
for $\Delta=1$ (3) GeV$^2$.  Below 1 GeV the prediction for ${\cal R}$ is much
more uncertain.  However, as discussed in Subsect.\ 2.6, we think that even
at $Q=0$ the result is reliable to within a factor of 2.  Conservatively, we
considered the effect of increasing the predicted ${\cal R}$ by a factor of 2
over the whole range, $0 < Q < 1$ GeV.  This affects the low-energy
$\bar{R}_{\scriptscriptstyle{PQW}}$ by 0.033 (0.011) for $\Delta=1$ (3)
GeV$^2$.  If we linearly add all the above-mentioned uncertainties we get a
total uncertainty of $\pm 0.07$ for $\Delta=1$ GeV$^2$ and $\pm 0.05$ for
$\Delta=3$ GeV$^2$.  Thus, the theoretical uncertainties are comparable to
the experimental uncertainties.

\subsection{Significance of the results}

    We can now discuss the significance of the agreement between theory
and experiment.  We first ask: How restrictive is the data?
To quantify the discussion we define a `straw-man' model for ${\cal R}$
in which ${\cal R}$ is the same as the OPT result down to 2 GeV, but then
follows the one-loop, 3-flavour form, $(12/27)(1/\ln Q^2/\Lambda_0^2)$, with
$\Lambda_0 \approx 0.2$ GeV, until it reaches a value $H$, at which it
remains frozen down to $Q=0$.  If the ``freeze-out'' value, $H$, is about 0.3,
then this `H model' is essentially equivalent to the OPT result.  If $H$
is much larger then this model gives a result for
$\bar{R}_{\scriptscriptstyle{PQW}}(0)$ that is too large
by more than the uncertainties just estimated.  We find that $H$'s above 2
are disfavoured by the data.  (As an illustration Fig.\ 8 shows the result
with $H = 4.6$, which is clearly ruled out.)  At the other extreme, the data
disfavour an $H$ less than 0.09.  Thus, although a wide range of $H$ values
can be tolerated, the data do imply that the couplant cannot grow very large
in the infrared region; nor can it remain too small.

    Next we ask: How predictive is the theory?  Because of the need for
smearing, the theory tells us almost nothing about the shape or structure
of the $e^+e^-$ data in the region below 1 GeV.  However, it does tell us
something about the average magnitude of the cross section.  The low-energy
data is, in fact, dominated by the $\rho$ peak.  After smearing with
$\Delta = 1 $ GeV$^2$, this contributes about 0.7 to
$\bar{R}_{\scriptscriptstyle{PQW}}$ below 1 GeV.  Thus a 10\% change in the
area under the $\rho$ peak would change $\bar{R}_{\scriptscriptstyle{PQW}}$
by the $\pm 0.07$ estimated uncertainty in the theoretical prediction.
We conclude that perturbative QCD can tell us, at least crudely, the size
of the $\rho$ resonance.

\subsection{The smeared derivative}

   As an extension of PQW's ideas we also considered a quantity:
\begin{equation}
\label{smder}
D(Q,\Delta) = \frac{2 \Delta}{\pi} \int_0^{\infty} ds^{\prime}
\frac{R_{e^+e^-}(\sqrt{s^{\prime}}) (s^{\prime}-Q^2)}
{ \{(s^{\prime}-Q^2)^2 +\Delta^2 \}^2 },
\end{equation}
which represents a ``smeared derivative'', in the sense that
\begin{equation}
\lim_{\Delta \to 0} D(Q,\Delta) = {\rm d} R_{e^+e^-} / {\rm d} Q^2.
\end{equation}
This provides a somewhat different test, though obviously not an independent
test, of the relationship between theory and experiment.  Its calculation
requires only straightforward modifications to the procedures used to
calculate $\bar{R}_{\scriptscriptstyle{PQW}}$.

   In Fig.\ 8 we compare the smeared derivatives from theory and
experiment for $\Delta = 2$  and 4 Gev$^2$.  For $\Delta = 2$ GeV$^2$ there
is good agreement at low energies, and the theory qualitatively gives the
first peak just below 3 GeV.  However, there is clearly insufficient smearing
in the charm region.  Increasing $\Delta$ to 4 GeV$^2$ greatly smooths out
both curves and gives quite good agreement.

\section{Phenomenological Virtues of a Frozen Couplant}

   The idea that the strong coupling constant, $\alpha_s(Q^2)$, ``freezes''
at low energies has long been a popular and successful phenomenological
hypothesis.  We first note that a freezing of $\alpha_s(Q^2)$
is a natural consequence of a picture where the gluon aquires an
effective, dynamical mass $m_g$ \cite{mg}.  Naively, this would
modify the leading-order, 3-flavour couplant to:
\begin{equation}
\label{mgform}
\frac{\alpha_s(Q^2)}{\pi} = \frac{12}{27} \,
\frac{1}{\ln[(Q^2 + 4m_g^2)/\Lambda_0^2]},
\end{equation}
a form that has been used in many phenomenological papers.  For $m_g$ a little
larger than $\Lambda_0$ this gives a zero-$Q$ value comparable to ours.  Note,
though, that the variation with $Q$ at low energies is somewhat different
from ours in Fig.\ 2.

  Another commonly used form is the ``hard-freeze'' form in which:
\begin{equation}
\label{Hform}
\frac{\alpha_s(Q^2)}{\pi} = \left\{
\begin{array}{ll}
 (12/27)(1/\ln Q^2/\Lambda_0^2), \quad &
{\mbox{\rm for}} \,\, Q^2 \ge Q_0^2, \\
 {\mbox{\rm constant}} \equiv H,  & {\mbox{\rm for}} \,\, Q^2 \le Q_0^2,
\end{array}
\right.
\end{equation}
with $H = (12/27)(1/\ln Q_0^2/\Lambda_0^2)$.  This is the ``$H$ model''
that we mentioned in Subsect.\ 3.5.  For $H \approx 0.26$ ({\it i.e.,}
$Q_0/\Lambda_0 \approx 2.3$) it is a close approximation to our
$\alpha_s(Q^2)/\pi$ shown in Fig.\ 2.

  We now briefly survey some of the phenomenological literature in order to
make two points:  (i) a frozen $\alpha_s$ provides a way to understand
many important facts in hadronic physics, and (ii) the values extracted
phenomenologically are very much in accord with our low-$Q$ value
$\alpha_s/\pi = 0.26$.  [Note that we quote $\alpha_s/\pi$ rather than
$\alpha_s$ values.]

  (a) {\it Total hadron-hadron cross sections}, although slowly rising at
very high energies, are remarkably constant over a wide energy range, and
their relative sizes correlate with their quark content in a very suggestive
way.  A simple and succesful description is provided by the 2-gluon-exchange
model \cite{gunion}, based on the Low-Nussinov model of the Pomeron
\cite{lownuss}.  This model requires a finite couplant at low momentum
transfer, and Ref.\ \cite{gunion} found a value $\alpha_s/\pi \approx 0.17$.
A recent version of this model, framed in terms of a dynamical gluon mass
($m_g = 0.37$ GeV, for $\Lambda_0 = 0.3$ GeV), is given in Ref.\ \cite{hkn}.
Another recent version of this model \cite{nik} uses the `H-model' form of
$\alpha_s(Q^2)/\pi$.  In order to fit the absolute magnitude of the
$\pi$-nucleon cross section, $Q_0$ needs to be about 0.44 GeV \cite{nik}
if $\Lambda_0 = 0.2$ GeV.  This corresponds to $H = 0.28$.  The same frozen
couplant has been used successfully in subsequent work on deriving nucleon
structure functions from the constituent-quark model \cite{barone}.

  (b) {\it Hadron spectroscopy} also points to a low-energy couplant of
around 0.2 -- 0.25 \cite{barnes}.  Godfrey and Isgur \cite{isgur} provide a
unified description of light- and heavy-meson properties in a ``relativized''
potential model with a universal one-gluon-exchange-plus-linear-confinement
potential.  For the model to work for light mesons it is crucial to
incorporate relativistic effects, and to employ a form of the couplant that
freezes at low energies.  Their fits yield a form of $\alpha_s(Q^2)/\pi$ that
freezes to about 0.19, and has a shape similar to ours.  In a fully
relativistic treatment Zhang and Koniuk \cite{zhang} can naturally explain
why the $\pi$ is so much lighter than the $\rho$.  The $\pi/\rho$ mass ratio
is a steeply falling function of the strong couplant, and the experimental
value occurs at $\alpha_s/\pi= 0.265$ \cite{zhang}.

  (c) {\it Hadron form factors} at low energies can be successfully treated
assuming a frozen couplant, as shown in Ref.\ \cite{ji}, which used the form
(\ref{mgform}) with $m_g \approx 0.1$ to 0.5 GeV, for $\Lambda_0 \approx 0.1$
GeV.

  (d) {\it Chiral soliton models of the nucleon} can fit a wide variety of
nucleon properties if one includes one-gluon exchange corrections with an
$\alpha_s/\pi$ of about 0.2 \cite{stern,duck}.  Ref.\ \cite{stern} finds that
the experimental deviation from the Gottfried sum rule, the $\Delta$-nucleon
mass difference, the first moment of the polarized proton structure function,
and the neutron-proton mass difference all require a common $\alpha_s/\pi$
value.  (However, the actual value found, 0.2, could be re-scaled by making a
different choice for another parameter in the model \cite{stern}.)  Other
nucleon properties are consistent with an $\alpha_s/\pi$ of this size
\cite{duck}.

  (e) {\it The $p_T$ spectrum in $W, Z$ production} in $pp$ or $p\bar{p}$
collisions can be successfully predicted by QCD right down to $p_T=0$
if multiple gluon radiation effects are appropriately re-summed \cite{WZpt}.
However, it is essential in the low-$p_T$ region to invoke a freezing of
$\alpha_s(p_T^2)$.  The form (\ref{mgform}) has been used, with
$4 m_g^2/\Lambda_0^2$ denoted by `$a$'.  Unfortunately, the results are very
insensitive to the parameter $a$; anything in the range 3 -- 100 gives an
acceptable fit to current data \cite{halzen,fletcher}.  This corresponds to a
range 0.1 -- 0.4 for the zero-$Q$ couplant.  Perhaps, future data will make
it possible to narrow this range.

  (f) {\it Jet properties} can be quite successfully described by the
``modified leading log approximation'' \cite{dok,khoze}, but to obtain
predictions at small momenta it is necessary to invoke a freezing of the
couplant.  Fits to data on heavy-quark-initiated jets give zero-$Q$ values
of $\alpha_s/\pi$ around 0.22 \cite{khoze}.  This value depends somewhat on
the form of $\alpha_s(Q^2)$ assumed in the fit, but it was found empirically
that the result for the integral:
\begin{equation}
\int_0^{1 {\mbox{{\small \rm GeV}}}} \! {\rm d} k \,
\frac{\alpha_s^{{\rm eff}}(k^2)}{\pi} \approx 0.2 \, {\mbox{\rm GeV}}
\end{equation}
was {\it fit invariant} \cite{khoze}.  Integrating our $\alpha_s/\pi$ in
Fig.\ 2 leads to precisely 0.2 GeV.

  (g) {\it Hadron-hadron scattering at very high energies} where the cross
sections rise asymptotically, but must satisfy unitarity, seems
to call for the `critical Pomeron' picture \cite{critpom}, at least as a
first approximation.  It seems that one could only hope to derive such a
picture from QCD if there is an infrared fixed point \cite{white}.  In fact,
White has argued for additional quarks, or colour-sextet quarks, in order
to have $N_f$ effectively equal to 16 \cite{white}.  (See the discussion
in Subsect.\ 2.6 above.)  However, our results imply that the infrared fixed
point persists down to low $N_f$.  This may mean that one can have all the
virtues of White's picture without the need for more quarks.

\section{Summary and Conclusions}

    We have applied OPT to the third-order QCD calculation of $R_{e^+e^-}$.
At energies above about 1 GeV there is every sign that the approximation
is healthy:  the perturbation series in the ``optimized'' scheme is well
behaved, and there is good agreement between second- and third-order results.
This was not true of the situation created by the old, incorrect $R_{e^+e^-}$
calculation \cite{old,max,oldopt} (see Table 2).  The contrast between the
`old' and `new' results emphasizes the point that the third-order $R_{e^+e^-}$
calculation provides a very real, empirical test of ``optimization'' ideas.
At the time, the statements \cite{oldopt} that the `old' third-order results
\cite{old} tended to cast doubt on the usefulness of ``optimization'' were
perfectly fair comment.  Because of this history we take especial satisfaction
in the transformed situation produced by the new result \cite{new}.

    Furthermore, contrary to the old situation, the optimized couplant now
shows a marked decrease from second to third order.  This is in accord with
the ``induced convergence'' conjecture that ``optimization'' naturally
cures the divergent-series problem \cite{optult,indcon}.

    The third-order OPT results remain finite down to $Q = 0$, with the
optimized couplant, $\alpha_s/\pi$, ``freezing'' to a value 0.26 below
300 MeV.  No {\it ad hoc} assumption was used to obtain this result: it is
the direct consequence of using the calculated $R_{e^+e^-}$ and $\beta$ series
coefficients as inputs to the ``optimization'' procedure specified in Ref.\
\cite{OPT}.

    It must be admitted that, at very low energies, the prediction
for ${\cal R}$ (the QCD correction term in $R_{e^+e^-}$) has a large
uncertainty.  Since third order is the lowest order at which it is even
possible to get finite infrared results, one should not be surprised if the
approximation is somewhat crude.  Nevertheless, as we discussed in Subsect.\
2.6, the qualitative conclusion that ${\cal R}$ remains small (say,
$0.3 \pm 0.3$ at $Q=0$) is inescapable in the context of OPT.

    The OPT prediction is supported by the data.  As we showed in Sect.\ 3,
the PQW-smeared $R_{e^+e^-}$ data is consistent with a perturbative QCD
description, provided that the couplant freezes to a modest value at low
energies.

    The hypothesis that the couplant freezes at low energies has been used
very successfully in a wide variety of phenomenological work, where the
low-energy couplant is treated as a free parameter to be fitted to experiment.
The values that emerge are quite comparable to ours.  There are some other
theoretical indications of a freezing of the couplant \cite{mg,gribov}, but
our evidence is remarkable in that it comes solely from perturbation theory
and RG invariance.  The predicted value, $\alpha_s/\pi = 0.26$, for the
frozen couplant is a purely theoretical number.  It does not depend on
knowing the value of $\Lambda_{\overline{\rm MS}}$, but only on knowing the
number of light quarks.

\newpage

\hspace*{-\parindent}{\bf Acknowledgements}

We thank A. Kataev and S. Larin for correspondence, and Ian Duck,
Robert Fletcher, Francis Halzen, Nathan Isgur, Valery Khoze, Chris Maxwell,
and John Ralston for helpful comments. \\
This work was supported in part by the U.S. Department of Energy under
Contract No. DE-FG05-92ER40717.

\subsection*{Appendix A: Numerical solution of the optimization equations}

    After expressing $\bar{r}_1$ and $\bar{r}_2$ in terms of $\bar{a}$ and
$\bar{c}_2$, using the $\rho_1$, $\rho_2$ definitions and (\ref{k3}), we have
to simultaneously solve the optimization equations, (\ref{eq1}, \ref{eq2}).
These define two curves in the $a$, $c_2$ plane whose intersection point we
seek.  In what we call the `spiralling' method, (\ref{eq2}) is first solved
for $\bar{a}$; then, with this $\bar{a}$, (\ref{eq1}) is solved for
$\bar{c}_2$; then, with this $\bar{c}_2$, (\ref{eq2}) is solved for $\bar{a}$;
and so on.  Which equation is solved for which variable is crucial; the
other choice would `spiral out' from the desired solution.  (The standard
``secant method'' \cite{nrecipe} was generally sufficient for solving the
individual optimization equations.)

    A convenient starting point for this iterative procedure was provided
by an approximate solution to the optimized equations due to Pennington,
Wrigley, and Mignaco and Roditi (PWMR) \cite{PWMR}.  This approximation
expands the optimization equations (\ref{eq1}, \ref{eq2}) as a series in
$\bar{a}$ and keeps only the lowest non-trivial term.  Noting that
$I = \bar{a}(1 - c \bar{a} + \ldots)$, it is easy to check that this gives:
\begin{equation}
\bar{r}_1 \approx 0 \quad\quad
\bar{r}_2 \approx -{\scriptstyle \frac{1}{3}} \bar{c}_2.
\end{equation}
One can improve this approximation by writing each $\bar{r}_i$ as a series
in $\bar{a}$ and successively equating coefficients of different orders in
$\bar{a}$ to zero.  To next order this gives:
\begin{equation}
\bar{r}_1 \approx {\scriptstyle \frac{1}{3}} \bar{c}_2 \bar{a}, \quad\quad
\bar{r}_2 \approx -{\scriptstyle \frac{1}{3}} \bar{c_2} +
{\scriptstyle \frac{1}{9}} c \bar{c_2} \bar{a}.
\end{equation}
[One may note that in $\bar{{\cal R}}^{(3)}$ there is a near cancellation
between the second and third order terms, $\bar{r}_1 \bar{a}$ and
$\bar{r}_2 \bar{a}^2$.  Thus, $\bar{{\cal R}}^{(3)}$ turns out to be closely
equal to $\bar{a}^{(3)}$.]

     The `spiralling' method worked well for $Q > 0.3$ GeV starting from the
PWMR solution.  However, at lower energies the PWMR approximation breaks down,
and does not provide a satisfactory initial guess.  In fact, at the infrared
fixed point one has instead, from (\ref{fpr}):
\begin{equation}
\bar{r}_1 = {\scriptstyle \frac{1}{2}} \bar{c}_2 \bar{a}^*, \quad\quad
\bar{r}_2 = -{\scriptstyle  \frac{2}{3}} \bar{c}_2.
\end{equation}
We therefore proceeded to low $Q$ in successive stages, utilizing the solution
at the previous $Q$ as the initial guess for the next lower $Q$.  We also
encountered a `creep' problem:  At low $Q$ the two curves representing the
``optimization'' equations become almost parallel (each being almost parallel
to the infrared boundary line $1 + c a + c_2 a^2 = 0$) and they cross at a
very small angle.  Thus, instead of `spiralling in' to the solution, one
creeps towards it stepwise.  The convergence is very slow and the danger is
that the solution can appear to have converged within the specified tolerance,
when in fact it still has a considerable way to go.  To avoid this pitfall
we would repeat the procedure from a different starting point, so as to creep
towards the solution from the other side.  In this way we could bracket the
true solution, and hence ensure reliable accuracy.

   We also tried the `intersection' method as an alternative.  Taking an
initial guess for $\bar{c}_2$, one solves for $\bar{a}$ in each of the two
optimized equations.  For each $\bar{a}$ one then solves the other equation
for $\bar{c}_2$.  This gives a pair of points on each of the two curves.
The straight lines that join up each pair should approximate the curves
themselves, and hence their intersection should approximate the desired
solution.  The procedure can then be iterated.  This method also worked well
for $Q > 0.3$ GeV starting from the PWMR solution.  At lower energies, where
the two curves become nearly parallel, this method did not suffer from the
`creep' problem, but it had the opposite vice:  it tended to make such a
large extrapolation in each iteration that it would become unstable and
erratic.

\subsection*{Appendix B: Flavour thresholds}

Since ${\cal R}$ has been calculated only with massless quarks, we are
really approximating ``full QCD'' with a {\it set} of effective theories,
each with a different number of massless quarks.
The $\Lambda_{\overline{\rm MS}}$ parameters of these theories need to be
appropriately matched, so that they correspond to a single, underlying
``full QCD'' theory.  The point is well explained by Marciano \cite{marciano},
who provides explicit formulas for matching $\Lambda_{\overline{\rm MS}}$
across thresholds.  Unfortunately his analysis uses a truncated expansion
of $a(\mu)$ in powers of $1/\ln(\mu/\Lambda)$, which would not be a valid
approximation at low energies; in particular at $s$-quark threshold.

   Our procedure was simply to require the optimized $\bar{{\cal R}}^{(3)}$
to be continuous at a threshold.  This was done numerically by running our
optimization program at the threshold energy ($Q = 2 m_q$) with both values
of $N_f$ and adjusting one of the $\Lambda_{\overline{\rm MS}}$ parameters
until the two $\bar{{\cal R}}^{(3)}$ results agreed.  Starting with
$\Lambda_{\overline{\rm MS}}^{(4)} = 230$ MeV for 4 flavours, we found
$\Lambda_{\overline{\rm MS}}^{(3)} = 281$ MeV, and
$\Lambda_{\overline{\rm MS}}^{(2)} = 255$ MeV.
[In terms of $\tilde{\Lambda}$ the corresponding values are:  257, 308, and
277 MeV for 4, 3, 2 flavours, respectively.]  Essentially the same results
were obtained if we required instead that $\bar{a}$ be continuous.  We checked
that this procedure agreed very closely with Marciano's formulas at both
$c$- and $b$-quark thresholds.

   It is noteworthy that we find $\Lambda_{\overline{\rm MS}}^{(2)}$ to be
smaller than $\Lambda_{\overline{\rm MS}}^{(3)}$, contrary to the pattern at
the higher thresholds.  Our final results are very insensitive to the
$\Lambda_{\overline{\rm MS}}^{(2)}$ value, however, because at energies below
$s$ threshold the $\bar{{\cal R}}$ results are essentially governed by the
infrared fixed point.

\clearpage

\begin{table}
\begin{center}
\begin{tabular}{c|c|c}
$N_f$ & $\rho_2$ & $\alpha_s^*/\pi$ \\
\hline
0     & $ -8.410$   & 0.313  \\
1     & $ -9.997$   & 0.280  \\
2     & $-10.911$   & 0.263  \\
3     & $-12.207$   & 0.244  \\
4     & $-13.910$   & 0.224  \\
5     & $-15.492$   & 0.208  \\
6     & $-17.665$   & 0.191  \\
\hline
\end{tabular}
\end{center}
\caption{$\rho_2$ invariants and fixed-point couplants for $N_f = 0$ to 6.}
\end{table}

\begin{table}
\begin{center}
\begin{tabular}{l||lccccc}
$N_f=5$, $Q=34$ GeV & Order & $\bar{a}$ & $\bar{r}_1$ & $\bar{r}_2$ &
$\bar{\cal R}$ & change \\ \hline
& & & & & & \\
 & 2nd & 0.0415 & $-0.599$ & --- & 0.0404 & --- \\
$\tilde{\Lambda}_{\overline{MS}} = 100$ MeV & 3rd & 0.0394 & $-0.301$ & 7.64 &
0.0394 & $-2.4$\% \\
($\Lambda_{\overline{MS}} = 87$ MeV) & old
& 0.0452 & \mbox{}+1.363 & $-29.48$ & 0.0453 & 12\% \\
& & & & & & \\
\hline
& & & & & & \\
 & 2nd & 0.0569 & $-0.588$ & --- & 0.0550 & --- \\
$\tilde{\Lambda}_{\overline{MS}}= 500$ MeV & 3rd & 0.0526 & $-0.405$ & 7.71 &
0.0526 & $-4.4$\% \\
($\Lambda_{\overline{MS}} = 436$ MeV)
& old & 0.0690 & \mbox{}+1.988 & $-27.59$ & 0.0694 & 26\% \\
& & & & & & \\
\hline
\end{tabular}
\end{center}
\caption{Comparison of second- and third-order optimized results:
`old' refers to third order with the old, incorrect result.}
\end{table}

\clearpage

\begin{table}
\begin{center}
\begin{tabular}{cc|ccccc}
$Q$ (GeV) & $N_f$ & $\bar{a}$ & $\bar{r}_1$ & $\bar{r}_2$ &
$\bar{\cal R}$ & error \\ \hline
& & & & & & \\
3.0 & 4 & 0.076 & $-0.53$ & 6.9  & 0.076 & 4\%   \\
1.0 & 3 & 0.126 & $-0.79$ & 6.3  & 0.126 & 10\%  \\
0.4 & 3 & 0.221 & $-1.77$ & 8.8  & 0.229 & 43\%  \\
0   & 2 & 0.263 & $-2.89$ & 14.6 & 0.330 & 100\% \\
& & & & & & \\
\hline
\end{tabular}
\end{center}
\caption{Illustrative third-order optimized results at low energies.
$\Lambda_{\overline{MS}} {\mbox{\rm (4 flavours)}} = 230$ MeV.  The
estimated fractional error is $|\bar{r}_2 \bar{a}^2|$.}
\end{table}

\clearpage

\section*{Figure Captions}

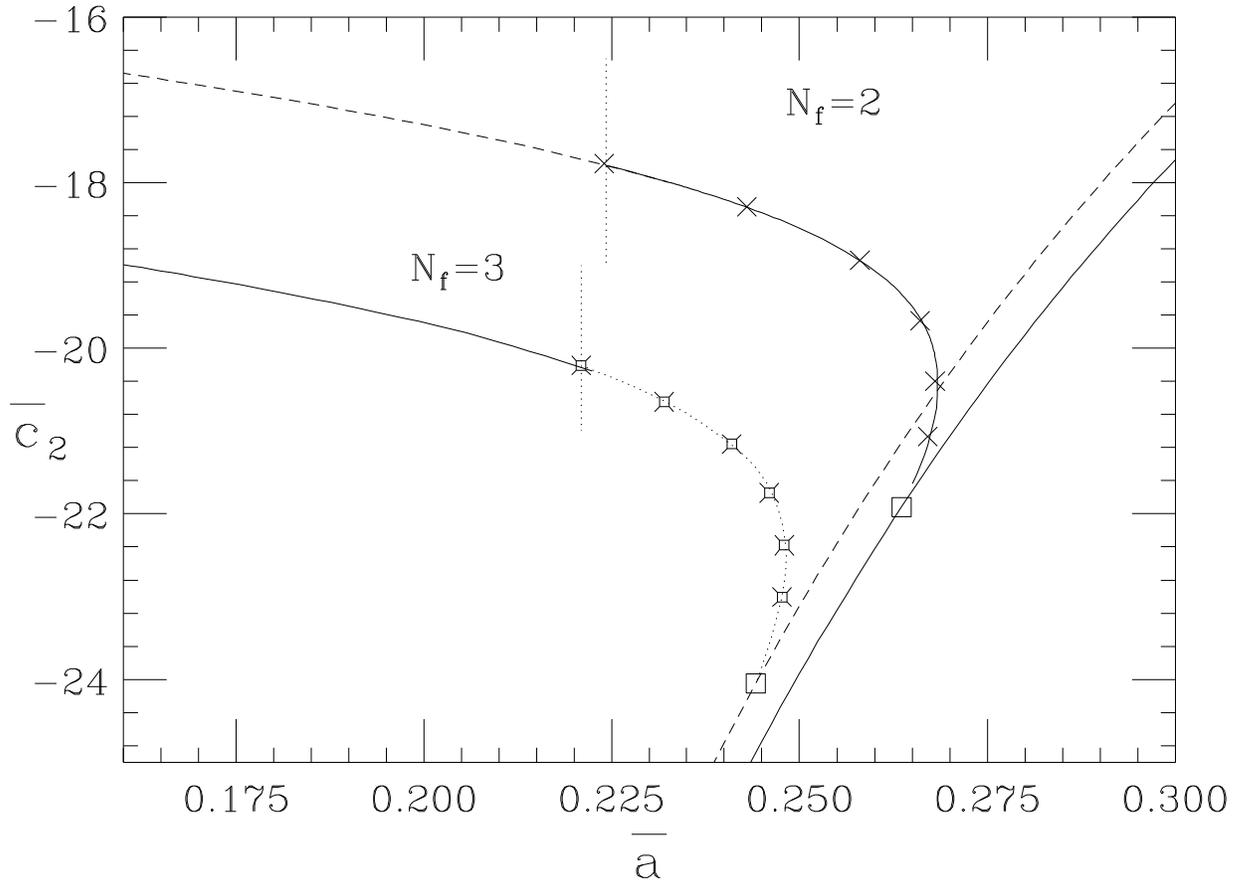
Fig.\ 1. The optimized solutions in the $a, c_2$ plane for 2 and 3 quark
flavours, in the low-energy region.  The open squares represent the
fixed-point solution, Eqs. (\ref{fpeq}, \ref{fpeqc2}), which lies on
the infrared boundary $1+ca+c_2 a^2 = 0$.  The boundary is shown by the
solid line ($N_f = 2$) or the dashed line ($N_f = 3$) at the right.  The
dotted vertical lines are to indicate $s\bar{s}$ threshold at $Q=0.40$ GeV
where $N_f$ changes from 3 to 2.  (The $\Lambda$ parameters are matched so
that $\bar{R}$ is continuous (see Appendix B.), but there are then slight
discontinuities in $\bar{a}$ and $\bar{c}_2$.)  The dotted line shows the
solution for a 3 flavor world down to $Q=0$, while the dashed line shows a
2 flavour world extending up towards 1 GeV.  The points shown are spaced at
0.05 GeV intervals from $Q=0.40$ GeV.

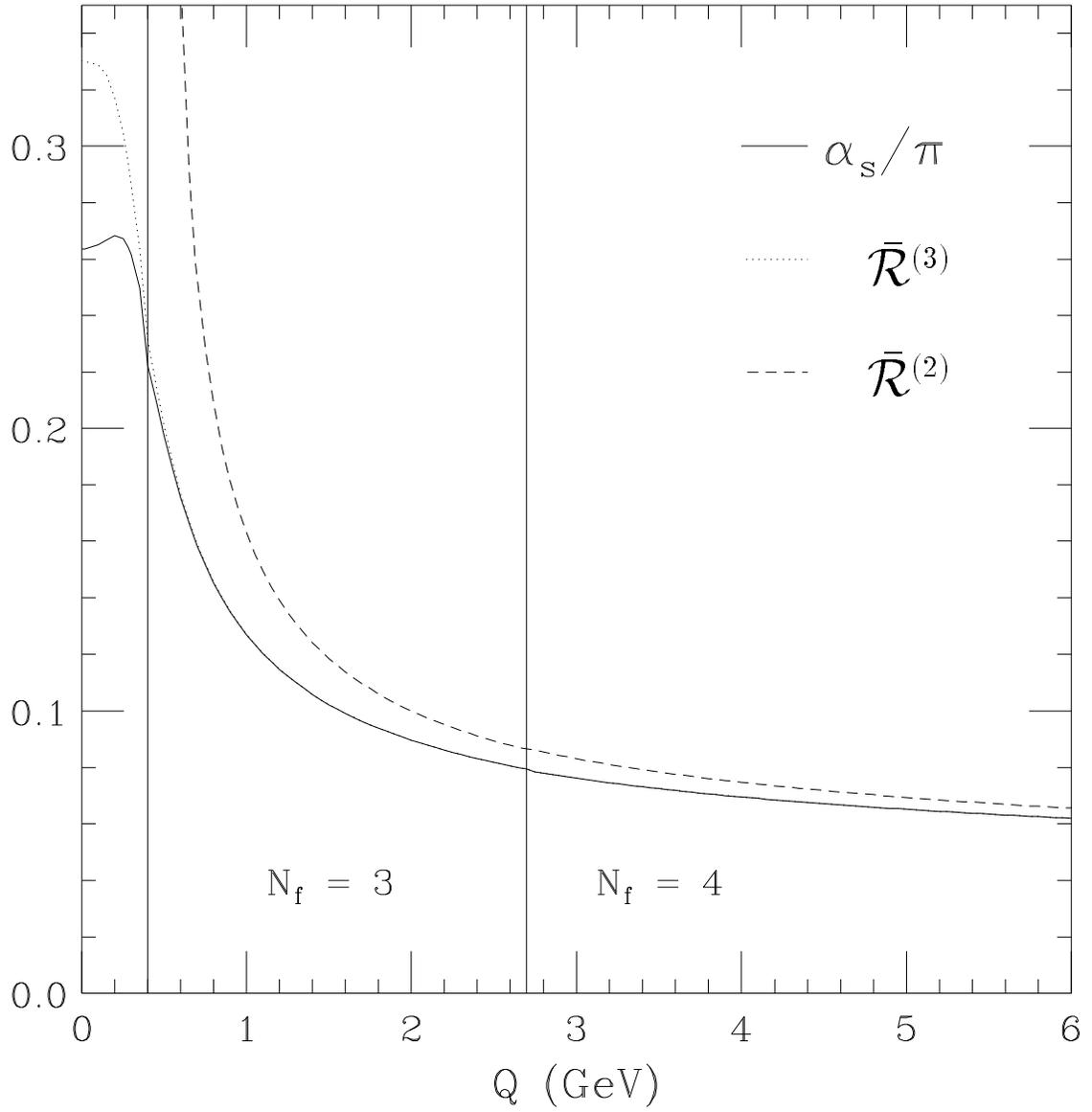
Fig.\ 2. The optimized third-order results for $\bar{a} = \alpha_s/\pi$
and $\bar{{\cal R}}^{(3)}$.  Also shown is the second-order result,
$\bar{{\cal R}}^{(2)}$.  Quark thresholds are indicated by the vertical
lines.

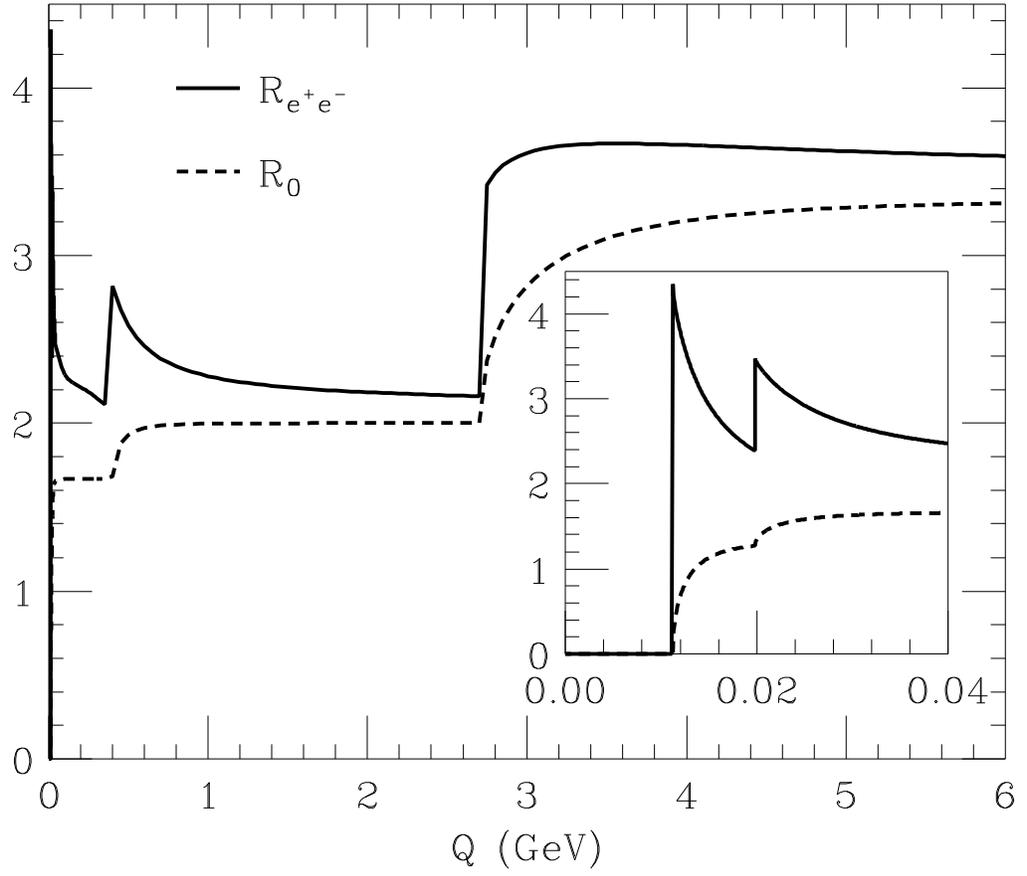
Fig.\ 3. The perturbative QCD prediction for $R_{e^+e^-}$ from third-order
OPT (solid line).  The inset shows the region around $u$ and $d$ quark
thresholds.  The dashed line is the parton-model prediction.

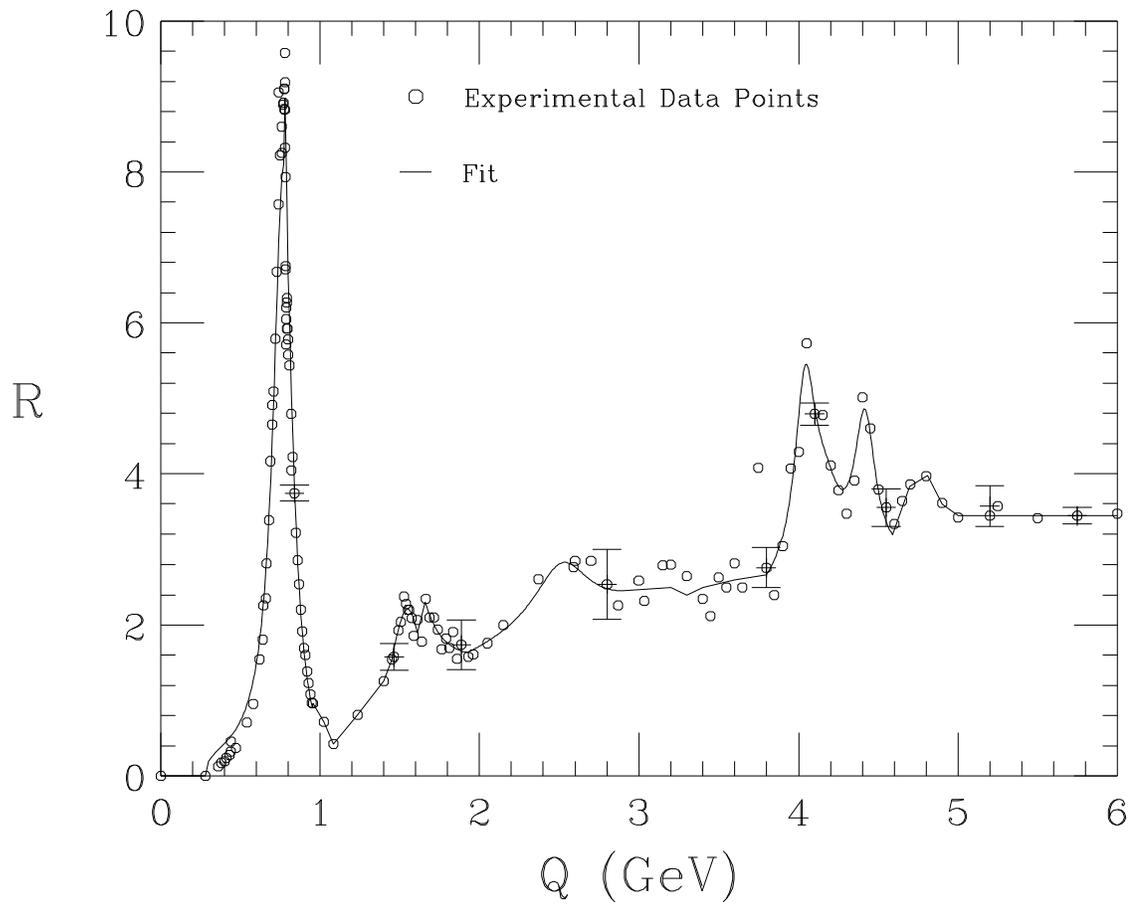
Fig.\ 4. Compilation of experimental $R_{e^+e^-}$ data (excluding narrow
resonances).  A few representative statistical error bars are shown.  The
solid line represents an `eyeball fit'.

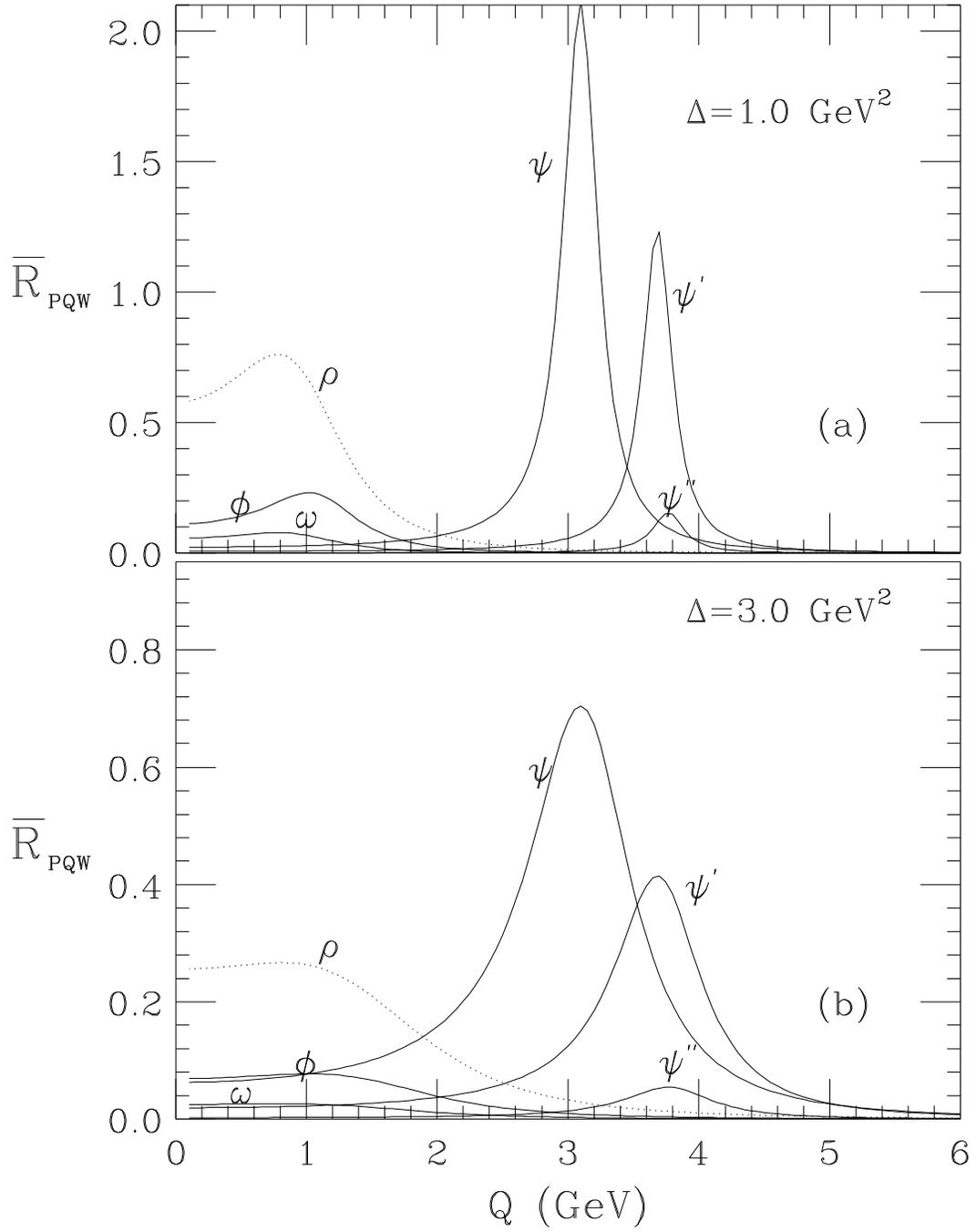
Fig.\ 5. The contributions of narrow resonances to
$\bar{R}_{\scriptscriptstyle{PQW}}$ for two values of the smearing parameter
$\Delta$.

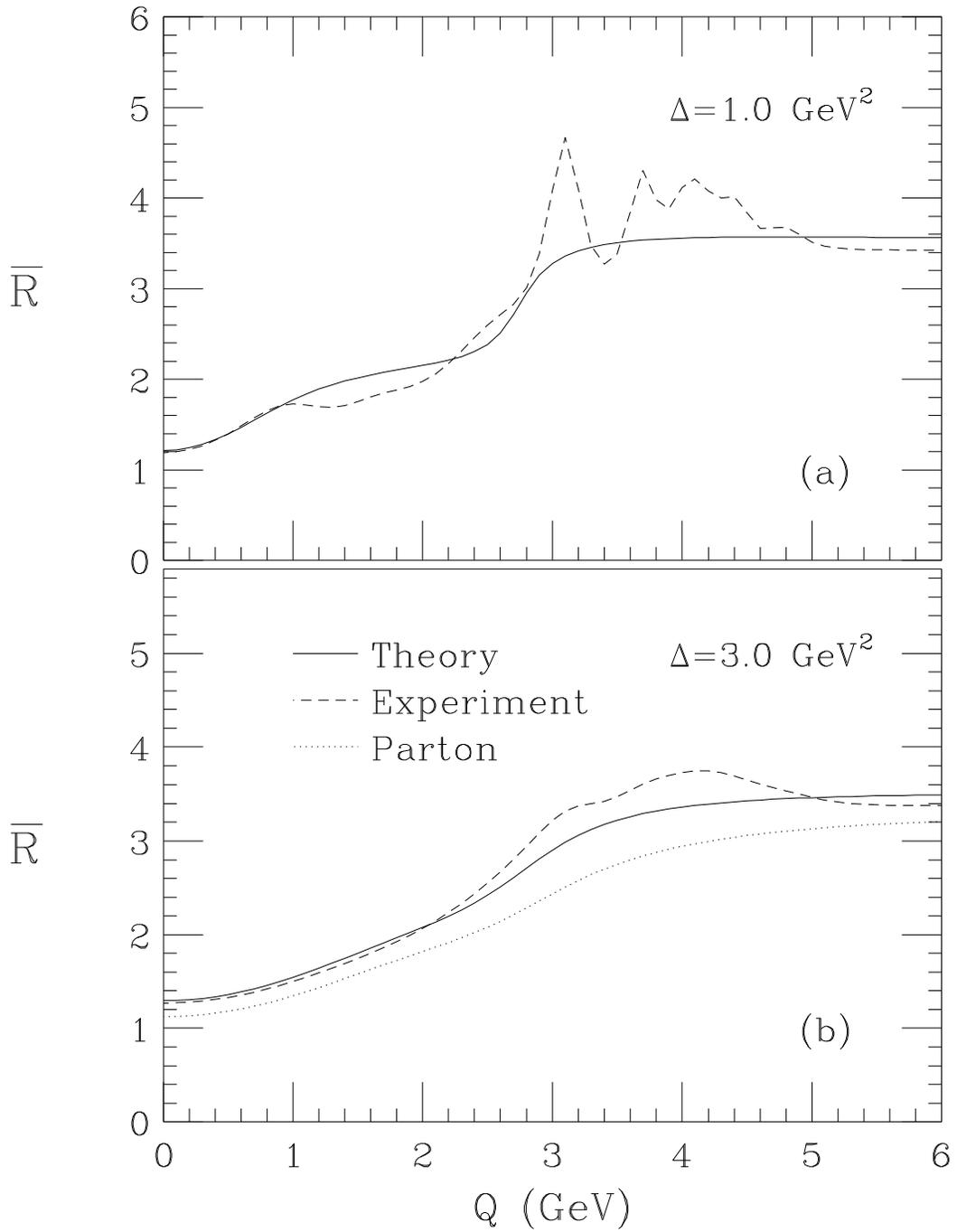
Fig.\ 6. Comparison of ``smeared'' theoretical and experimental results.
The parton-model result is shown by the dotted line in (b).

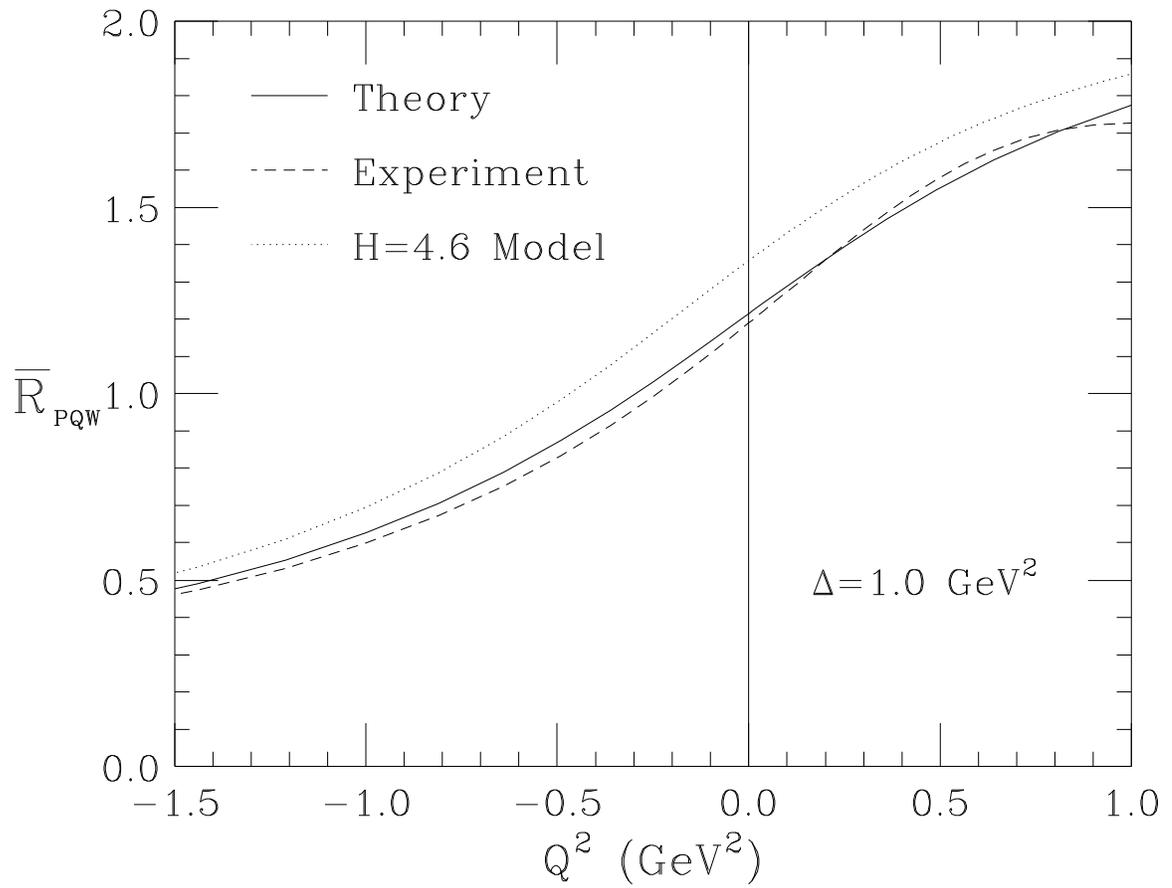
Fig.\ 7. Comparison of ``smeared'' results extended to spacelike $Q^2$.
The dotted line shows a `straw-man' model in which the couplant becomes
large at low energies (see subsect.\ 3.5).

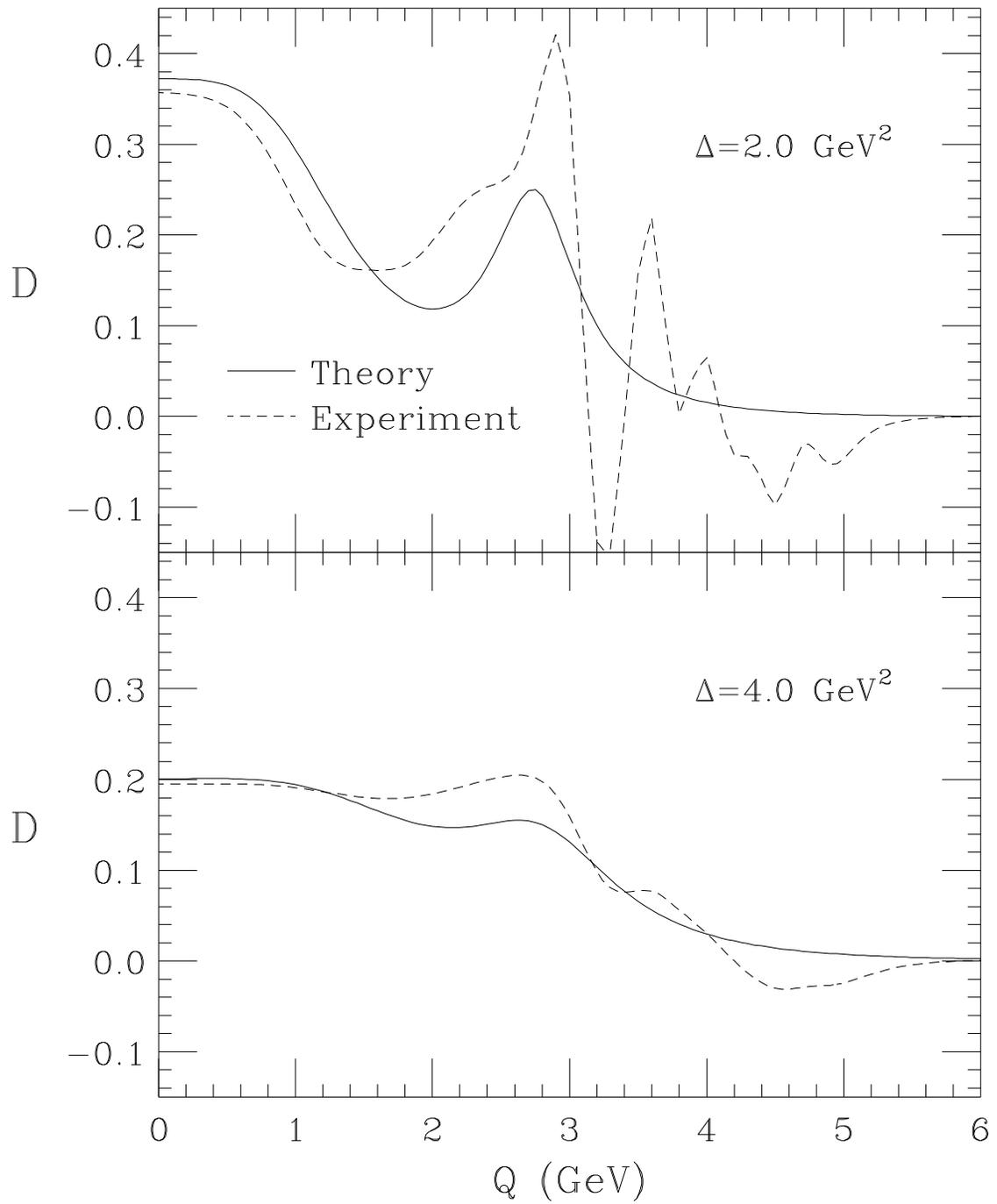
Fig.\ 8. Comparison of theoretical and experimental results for the
``smeared derivative'' (Eq. (\ref{smder})) for two values of $\Delta$.


\begin{thebibliography} {99}
\bibitem{fncg}
  Two schemes, which for the same $\mu$ value are related by
  $a^{\prime} = a(1 + v_1 a + \ldots)$, have $\tilde{\Lambda}$ parameters
  which are related by
  $\ln(\tilde{\Lambda}^{\prime}/\tilde{\Lambda}) = v_1/b$.  This relation
  is {\it exact} \cite{CG,OPT}.
\bibitem{fac}
  Fixed-point behaviour will also be found in RS's which are sufficiently
  close to the ``optimized'' scheme.  This is the case, for instance, in the
  `FAC' or ``effective charge'' scheme (the RS in which $r_1=r_2=0$).  However,
  this scheme gives an $a^*$ that is 64\% larger than ours, for $N_f=2$.
\bibitem{fnrep}
  Obviously, the procedure may be repeated with different trial values of
  $\Lambda_{\overline{\rm MS}}$ if one seeks to determine the value that
  gives the best fit to the $R_{e^+e^-}$ data.
\end{thebibliography}
\end{document}